\documentclass[12pt,a4paper,final]{iopart}

\usepackage[immediate]{silence}
\usepackage{iopams}  
\usepackage[final]{graphicx}
\usepackage[breaklinks=true,colorlinks=true,linkcolor=blue,urlcolor=blue,citecolor=blue]{hyperref}

\usepackage{placeins}

\usepackage{doi}

\usepackage{multirow}

\usepackage{blindtext}

\usepackage{notes2bib}

\expandafter\let\csname equation*\endcsname\relax
\expandafter\let\csname endequation*\endcsname\relax
\usepackage{amsmath}

\usepackage{chemformula}

\usepackage{microtype}

\usepackage{subcaption}

\usepackage[forbid-literal-units=true]{siunitx}
\usepackage[obeyFinal]{todonotes}

\usepackage{showlabels}

\usepackage{booktabs}

\usepackage[perpage]{footmisc}

\usepackage{cite}

\usepackage{braket}

\usepackage[capitalize,noabbrev]{cleveref}
\creflabelformat{equation}{#2#1#3}

\usepackage{mathtools}

\usepackage[raggedright]{titlesec}
\titleformat{\subsubsection}[hang]{\normalfont\normalsize\bfseries}{\thetitle}{1em}{}
\titlespacing*{\subsubsection}{0pt}{3.25ex plus 1ex minus .2ex}{0.5em}

\usepackage{mathtools,bm}
\usepackage{letltxmacro}

\LetLtxMacro\oldvec\vec
\RenewDocumentCommand \vec { o m o }
{
	\IfNoValueTF {#1}
		{
			\IfNoValueTF {#3}
				{  {  \oldvec{\mathrm{#2}} }	                }
				{  {  \oldvec{\mathrm{#2}} }_{\mathrm{#3}} }
		}
		{
			\IfNoValueTF {#3}
				{ \mathrm{ {#2}_{#1}		} }
				{ \mathrm{ {#2}_{#3,#1} 	} }
		}
}

\RenewDocumentCommand \mat { o m o }
{
	\IfNoValueTF {#1}
		{
			\IfNoValueTF {#3}
				{ { { {{#2}} }	        } }
				{ { { {{#2}} }_{{#3}} } }
		}
		{
			\IfNoValueTF {#3}
				{ { { {#2}_{#1} }		} }
				{ { { {#2}_{#3,#1} }  	} }
		}
}

\NewDocumentCommand \unit { o m o }
{
	\IfNoValueTF {#1}
		{
			\IfNoValueTF {#3}
				{  {\hat{ {#2} 	 }} }
				{   \hat{{#2}}_{#3} }
		}
		{
			\IfNoValueTF {#3}
				{ \mathrm{ {#2}_{#1}		} }
				{ \mathrm{ {#2}_{#3,#1} 	} }
		}
}

\newcommand{\sm}{\SI{7}{\micro\meter}}
\newcommand{\smh}{\num{7}-\si{\micro\meter}}
\newcommand{\B}[1]{\ensuremath{ \vec{B}[#1] }}
\newcommand{\Yb}{\ch{^{171}Yb+}}

\begin{document}

\title{Measurement of differential polarizabilities at a mid-infrared wavelength in \Yb{}}

\author{C~F~A~Baynham$^{1,2}\footnote{Corresponding author}$, E~A~Curtis$^{1}$, R~M~Godun$^{1}$, J~M~Jones$^{1,3}$, P~B~R~Nisbet-Jones$^{1}$, P~E~G~Baird$^{2}$, K~Bongs$^{3}$, P~Gill$^{1,2}$}
\address{$^1$National Physical Laboratory, Hampton Road, Teddington TW11 0LW, United Kingdom}
\address{$^2$Clarendon Laboratory, University of Oxford, Parks Road, Oxford OX1 3PU, United Kingdom}
\address{$^3$School of Physics and Astronomy, University of Birmingham, Birmingham B15 2TT, United Kingdom}

\author{T~Fordell, T~Hieta, T~Lindvall}
\address{VTT Technical Research Centre of Finland Ltd, Centre for Metrology MIKES, P.O.Box 1000, FI-02044 VTT, Finland}

\author{M~T~Spidell, J~H~Lehman}
\address{National Institute of Standards and Technology, 325 Broadway, Boulder, CO 80504, United States}

\ead{charles.baynham@npl.co.uk}

\begin{abstract}

An atom exposed to an electric field will experience Stark shifts of its internal energy levels, proportional to their polarizabilities. 
In optical frequency metrology, the Stark shift due to background black-body radiation (BBR) modifies the frequency of the optical clock transition, and often represents a large contribution to a clock's uncertainty budget. For clocks based on singly-ionized ytterbium, the ion's complex structure makes this shift difficult to calculate theoretically. 
We present a measurement of the differential polarizabilities of two ultra-narrow optical clock transitions present in \Yb{}, performed by exposing the ion to an oscillating electric field at a wavelength in the region of room temperature BBR spectra. By measuring the frequency shift to the transitions caused by a laser at $\lambda=\SI{7.17}{\micro\meter}$, we obtain values for scalar and tensor differential polarizabilities with uncertainties at the percent level for both the electric quadrupole and octupole transitions at \SI{436}{\nano\meter} and \SI{467}{\nano\meter} respectively. 
These values agree with previously reported experimental measurements and, in the case of the electric quadrupole transition, allow a 5-fold improvement in the determination of the room-temperature BBR shift. 

However, we note significant concerns over the validity of the uncertainty charactarization presented and draw the reader's attention to the Note on applicability section for a discussion. 
\end{abstract}

% \vspace{-0.5cm}
% \pacs{32.10.Dk}
% \vspace{3mm}

% % Keywords required only for MST, PB, PMB, PM, JOA, JOB? 
% \noindent{\it Keywords}: Article preparation, IOP journals
% % Uncomment for Submitted to journal title message
% \submitto{\NJP}

%%%%%%%%%%%%%%%%%%%%%%%%%%%%%%%%%%%%%%%%%%%%%%%%%%
% Remove this section when ready for submission  %
%%%%%%%%%%%%%%%%%%%%%%%%%%%%%%%%%%%%%%%%%%%%%%%%%%

% Get current git version
% Note: this command requires a) git installed and b) shell escape in latex enabled
% \immediate\write18{git describe --always --dirty --long --abbrev=6 > \jobname_desc.nogit.txt }
% \fbox{Manuscript version: \textit{\input{\jobname_desc.nogit.txt}}}

\maketitle

\section*{Note on applicability}

In \cref{sec:absolute_polarizabilities}, this paper presents an experiment to determine the absolute value of the differential polarizability of two clock transitions in \Yb{} at $\lambda=\sm{}$, which is then used in \cref{sec:deducing_bbr_frequency_shift} to deduce the differential polarizabilities at dc.
The method relies upon integrating atomic frequency shifts across the whole profile of an incident laser beam at \sm{}: if any part of the beam's profile lies outside the region sampled, the deduced ``sum of shifts" over the beam would be incorrect, resulting in an underestimate of the absolute polarizability.
\Cref{para:beam_capture} presents an attempt to quantify the measurement uncertainties, however it simply cannot be proven that there was no additional laser power outside the region that was sampled.

For this reason, the authors cannot demonstrate full confidence in the uncertainty analysis presented in \cref{tab:uncertainty_budget} and in the dependent results in \cref{sec:deducing_bbr_frequency_shift,tab:final_results}, and will therefore not pursue the publication of this manuscript. 

\section{Introduction} % (fold)
\label{sec:introduction}

Singly-ionized \Yb{} has two ultra-narrow optical frequency transitions that have been selected as secondary representations of the SI second: an electric quadrupole (E2) transition and an electric octupole (E3) transition, respectively $\ch{4 f^{14} \; 6 s \; ^2S_{1/2}}(F=0, M_F=0) \to \ch{^2D_{3/2}}(F=2, M_F=0)$ and $\ch{^2S_{1/2}}(F=0, M_F=0) \to \ch{^2F_{7/2}}(F=3, M_F=0)$~\cite{Roberts1997,Hosaka2009,Huntemann2016,King2012,Godun2014}. 
When realising a frequency standard based on these transitions, all sources of frequency shift that can perturb the natural atomic frequency must be characterized and either prevented or subtracted. This includes the frequency shift arising from the ion's interaction with the electric field in the thermal background, i.e.\ the Stark shift due to environmental black-body radiation (BBR).

In the presence of an oscillating electric field $\vec{E}$, an energy level $\gamma$ with quantum numbers $J$ and $F$ in an atom is shifted by an amount $\delta W$, given by 
\begin{alignat}{2}	% For aligning multi-line equations
\nonumber % Don't number first line
\delta W(\gamma, J, F, \vec{E}, \theta)
&=&\quad -&\frac{1}{2}\alpha(\gamma, J, F, \theta) \; \langle \vec{E}^2 \rangle \\
&=&  -&\frac{1}{2} \left[ \alpha^{SC}(\gamma, J) + \frac{3M_F^2 - F(F+1)}{2F(2F-1)} (3\cos^2\theta - 1) \; \alpha^{TEN}(\gamma, J, F) \right] \langle \vec{E}^2 \rangle
\,,
\label{eq:shift}
\end{alignat}
where $\theta$ is the angle between the electric field and the atom's quantisation axis; $\alpha$~is the polarizability of the state and $\alpha^{SC}, \alpha^{TEN}$ are the polarizability's scalar and tensor components~\cite{Itano2000}. Note that the vector component of the polarizability is not present since both clock transitions involve states with $M_F = 0 \to 0$~\cite{Manakov1986,LeKien2013}.

\begin{figure}[tb]
	\centering
	\includegraphics[width=\textwidth]{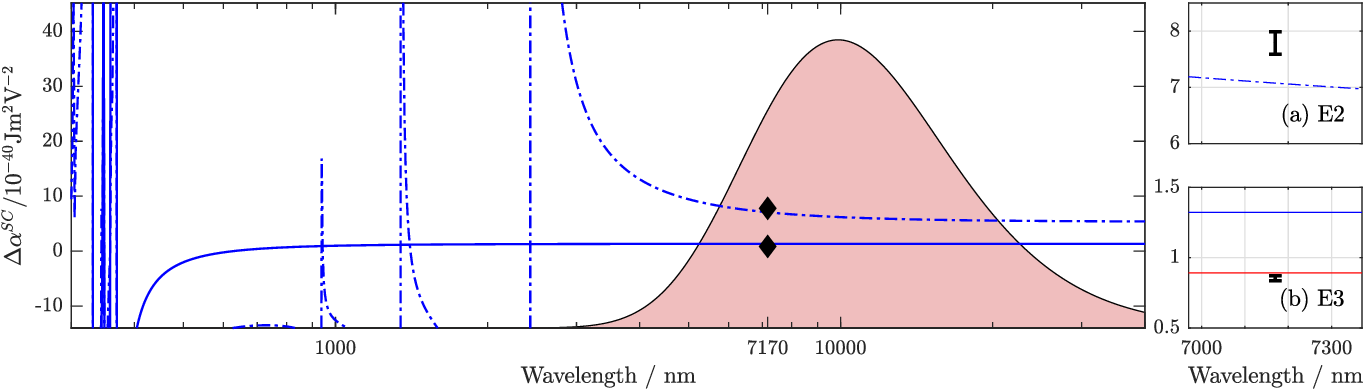}
	\caption[Differential polarizabilities of the Yb+ ion]{\textit{Main plot} The solid and dot-dashed lines show the differential polarizabilities of the E3 and E2 transitions respectively, calculated with oscillator strength corrected theory (see \cite{Biemont1998,Lea2006}). The solid fill shows the spectrum of black-body radiation at \SI{298}{\kelvin} in arbitrary units. The points show the values experimentally measured in this work. \\
	\textit{Sub plots} Close-up views of the (a) E2 and (b) E3 differential polarizabilities, where the blue lines are those from the main plot, the red line results from a least-squares fit to measurements taken in the near-IR for the E3 transition (calculated and reported in~\cite{Huntemann2016}) and the points show the measurements taken in this work at $\lambda = \SI{7.17}{\micro\meter}$. }
	\label{fig:polarizability-and-BBR}
\end{figure}

This perturbation to the atomic energy levels results in an observed frequency shift $\delta\nu$ given by
\begin{align}
	h \, \delta\nu(\vec{E}, \theta) 
		&= \delta W_2 - \delta W_1 		\nonumber \\
		&= - \frac{1}{2} \Delta\alpha_{21}(\theta) \; \langle \vec{E}^2 \rangle
		\,,
		\label{eq:simple_polar}
\end{align}
where $\Delta \alpha_{21} = \alpha_2 - \alpha_1$ is the differential polarizability of the transition between states 1 and 2 and is independent of the magnitude of the electric field. 
%For two measurements with the same intensity of light at the ion, the ratio of their induced frequency shifts is therefore equal to the ratio of their differential polarizabilities. 
The differential polarizability of a transition is frequency dependent but, for electric fields in the frequency range typical of room temperature BBR spectra, the two clock transitions' differential polarizabilities ($\Delta\alpha_{E2}$ and $\Delta\alpha_{E3}$) are smoothly varying and tend asymptotically to their dc values, as shown in \cref{fig:polarizability-and-BBR}~\cite{Biemont1998}.

The complex structure of the ytterbium ion makes theoretical calculations of these polarizabilities difficult~\cite{Lea2006,Roy2017} and, unlike in nearly temperature-insensitive clock species such as lutetium~\cite{Arnold2017}, the room-temperature BBR shift is large enough to make the accurate determination of these polarizabilities vital for clock operation with fractional frequency uncertainty at the $10^{-18}$ level. In this work, we therefore present direct measurements of the frequency shifts of the E2 and E3 transitions in \Yb{}, induced by the electric field of a laser with a mid-IR wavelength of \SI{7.17}{\micro\metre}, representative of a typical room-temperature thermal environment. From this we deduce values for the differential polarizabilities of both transitions that are of direct use in determining the BBR shifts for a \Yb{} frequency standard. 

% section introduction (end)

\section{Experimental method} % (fold)
\label{sec:experimental_method}

\begin{figure}[tb]
	\centering
	\includegraphics[width=\textwidth]{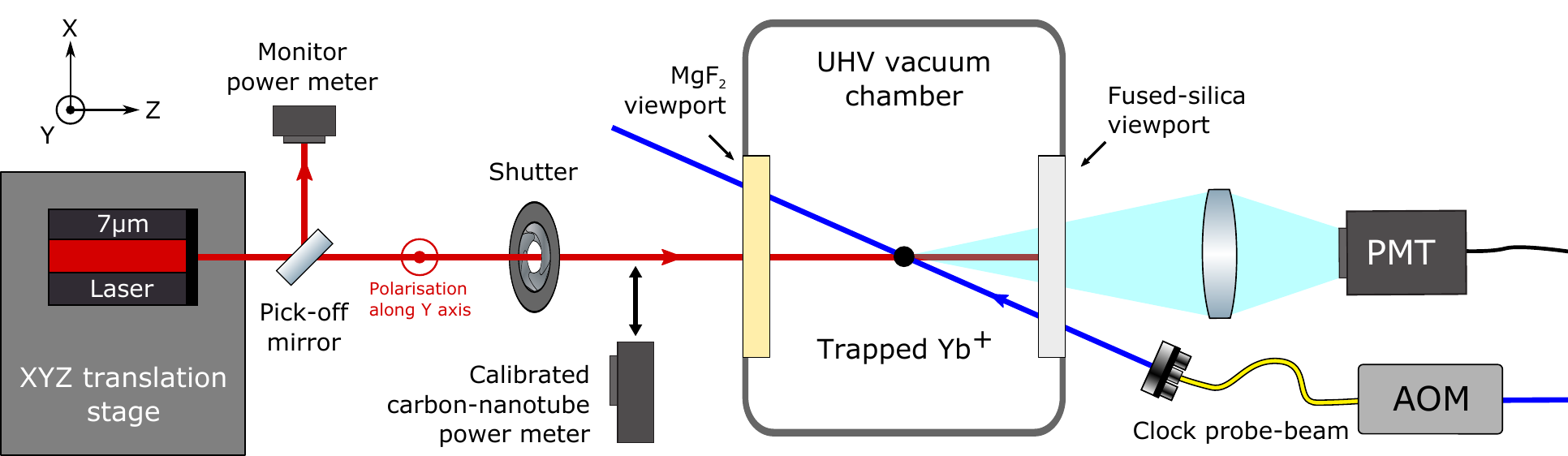}
	\caption{A trapped \Yb{} ion is interrogated by an ultra-stable probe laser at either \SI{436}{\nano\meter} or \SI{467}{\nano\meter}. The transition frequency is perturbed by the electric field created by an incident laser beam of $\lambda=\SI{7.17}{\um}$. The laser is mounted on a computer-controlled translation stage to allow the full intensity profile of the beam to be explored. The \smh{} beam is inclined by \SI{4}{\degree} clockwise in the horizontal plane with respect to the z axis (which is perpendicular to the vacuum chamber window), with polarisation along the y axis. 
	Total power in the beam is continuously monitored for stability, and can be measured with a calibrated carbon nanotube-based detector. }
	\label{fig:experimental_setup}
\end{figure}

In our experimental setup, as shown in \cref{fig:experimental_setup}, a single ion of \Yb{} is trapped in an rf Paul trap of the end-cap style, described exactly by~\cite{Nisbet-Jones2016}. For this experiment, the trap was operated with a confining potential oscillating at \SI{13}{\mega\Hz}, producing secular modes of frequency \SIlist{438;462;857}{\kilo\hertz}. 
Two ultra-stable lasers at \SI{436}{\nano\meter} and \SI{467}{\nano\meter} are used to drive the ultra-narrow quadrupole (E2) and octupole (E3) optical clock transitions (for a description of the operation of this optical clock see~\cite{King2012,Godun2014}). 
To excite the two clock transitions, a Rabi excitation probe scheme was used. For the E3 and E2 transitions respectively the locking routine used \SI{100}{\ms} probe times with a \SI{62}{\percent} duty cycle or \SI{30}{\ms} probe times with a \SI{45}{\percent} duty cycle.

In addition to this, a quantum cascade laser (QCL)
\bibnote{Hamamatsu LE0088 with output power $\approx \SI{20}{\milli\watt}$. Diode current supplied by an ILX Lightwave LDX-3232 current controller. The LE0088 is a custom laser model, most similar to the range of diodes given at \url{https://www.hamamatsu.com/jp/en/product/category/1001/1006/1021/index.html} but with an integrated focussing optic as in the range at \url{https://www.hamamatsu.com/jp/en/product/category/1001/1006/1027/index.html}. }
oscillating with  $\lambda = \SI{7.17}{\micro\meter}$ was mounted to a NanoMover translation stage%
\bibnote{NanoMover 11NCM001 actuators with Nanomotion II Model 11NCS101 driver unit by Melles-Griot}
and aligned onto the ion almost along the z-axis, offset by \SI{4}{\degree} to prevent any reflection from the rear window interacting with the ion. The vacuum chamber of the system was fitted with a window made from \ch{MgF2}, chosen for its transmissivity at this wavelength~\cite{Nisbet-Jones2016} (transmissivity requirements for other frequencies used in our experiment, particularly UV light at \SI{369}{\nano\meter}, precluded use of e.g.\ a \ch{ZnSe} window whose transmissivity at \sm{} would be slightly higher). The laser was linearly polarised with electric field always in the y-direction. % The spot-size at the ion was measured to be \SI{100}{\micro\meter}\todo{spot size}. 

\begin{figure}[tb]
	\centering
	\includegraphics[width=\textwidth]{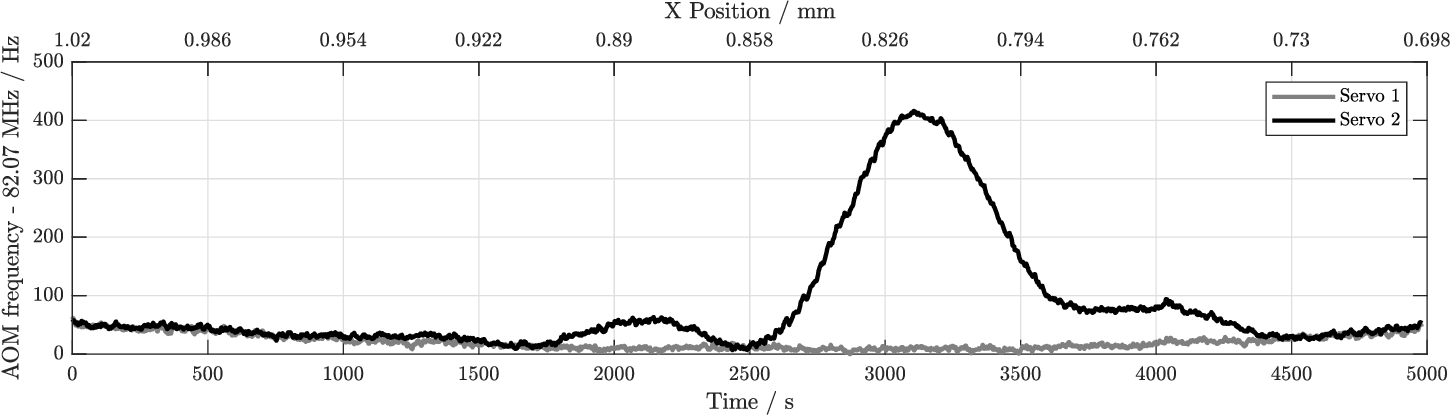} 
	\caption{Typical data taken as the \smh{} laser beam is translated over the ion's position (upper axis). Servo 1 shows the clock laser frequency in the absence of perturbation, while servo 2 records the Stark shift induced by the \smh{} radiation.}
	\label{fig:typical_data}
\end{figure}

The \smh{} laser's electric field induces a frequency shift of the clock transition as described by \cref{eq:simple_polar}. This shift was measured by running two interleaved frequency locks to the atomic transition: one with the \smh{} light incident on the ion and another with it blocked by a shutter.
All other experimental conditions were kept identical between the two servos so we attribute any observed frequency difference as entirely due to the Stark shift of the infrared light.
\cref{fig:typical_data} shows data taken in this arrangement, as the translation stage was moved to scan the position of the \smh{} laser beam over the ion. The laser was typically in a given position for a period of \SI{30}{\second} to allow frequency data to be acquired, before stepping to the next position. 

Control of the ion's quantisation axis was achieved by the application of an external magnetic field to set or vary the angle $\theta$ in \cref{eq:shift}. 

% section sec:experimental_method (end)

\section{Polarizability ratios} % (fold)
\label{sec:polarizability_ratios}

Without characterising the absolute strength of the electric field at \sm{}, information can already be obtained about the relative polarizabilities between different transitions or between different orientations of the ion, by making measurements in a constant electric field.  For these measurements the intensity maximum of the \smh{} beam was positioned onto the ion by maximizing shifts to the clock transitions, increasing our signal-to-noise ratio, while also minimising sensitivity to small excursions in the beam position.  

By monitoring the power at the beam pick-off shown in \cref{fig:experimental_setup}, we determined that the total power of IR light produced by the \smh{} laser in the plane of the ion was stable at the $\delta P / P \leq 6\times10^{-5}$ level over time-scales of $10^3$ to $10^4$ seconds. Note that this stability refers to the total power in the beam, not the intensity experienced by the ion which is subject to further uncertainties such as beam pointing which will be addressed in \cref{sec:absolute_polarizabilities}. We attribute this surprisingly high power stability to the inherent stability of QCLs\cite{Mezzapesa2013} and note that even if this stability were to be degraded by a factor of 100 this would not affect the results presented here. 

We will first discuss the ratio of the differential scalar polarizabilities between the E3 and E2 clock transitions, $\Delta\alpha^{SC}_{E3} / \Delta\alpha^{SC}_{E2}$. The frequency shifts from the incident \smh{} light were measured for each clock transition in turn, in a set of orthogonal magnetic field directions 
chosen to simultaneously maximise the three Rabi frequencies for the clock probe-beam's fixed direction. 
The average of the frequency shifts across the three orthogonal orientations of the ion revealed the contribution from the differential scalar polarizability component for each transition, with the tensor component averaging to zero over the three orthogonal directions \cite{Itano2000}.  Taking the ratio of these average frequency shifts for the two transitions, recorded in equal \smh{} electric field strengths, gives the ratio $\Delta\alpha^{SC}_{E3} / \Delta\alpha^{SC}_{E2}$, as shown in \cref{tab:ratios}.  Evaluating the uncertainty required consideration of both statistical uncertainties in the measured frequency shifts, and also non-orthogonalities in the applied magnetic field directions.  Both these uncertainties dominated over any changes in the field strength or direction of the \smh{} electric field during the measurements.  To evaluate the uncertainty due to non-orthogonality of the magnetic fields, a simulation was performed based on measurements of the field strength taken via the Zeeman components of the E2 transition. For the details of this method, see \ref{appendix:magnetic_field_orthogonality}.

% \begin{table}[tb]
% \centering

% \begin{tabular}{@{}lrrrr@{}}
% \toprule
% %				& Octupole					& Quadrupole 							&& \\
% 				& $\Delta\alpha^{SC}_{E3} / \Delta\alpha^{TEN}_{E3}$ 	& $\Delta\alpha^{SC}_{E2} / \Delta\alpha^{TEN}_{E2}$ 
% 				& \phantom{xx} % This "phantom" entry is invisible and used only for spacing
% 																		& $\Delta\alpha^{SC}_{E3} / \Delta\alpha^{SC}_{E2} $ \\
				
% 				\cmidrule{2-3} \cmidrule{5-5}

% Ratio 			& ? 	& ?								&& ?					\\
% Uncertainty 	& ? 	& ?								&& ?					\\ \bottomrule
% \end{tabular}

% \caption{Observed ratios between the scalar and tensor differential polarizabilities of the two clock transitions in \ch{Yb+}. E2 and E3 refer to the electric quadrupole and octupole clock transitions in \ch{Yb+} respectively.}
% \label{tab:ratios}

% \end{table}

\begin{table}[tb]
\centering

% Data for this table comes from the Excel file Scalar_Tensor shifts.xlsx, as saved in this repo

\begin{tabular}{llr}
\toprule
Quantity                                           &                                                      &  Ratio      \\ \midrule
\multicolumn{2}{l}{{Quadrupole : Octupole}}        &                                                      \\
\phantom{xxxx}                                     & $\Delta\alpha^{SC}_{E2} / \Delta\alpha^{SC}_{E3}$    &  9.13(14)   \\
\multicolumn{2}{l}{{Scalar : Tensor}}              &                                                      \\
                                                   & $\Delta\alpha^{SC}_{E3} / \Delta\alpha^{TEN}_{E3}$   &  -4.13(10)  \\
                                                   & $\Delta\alpha^{SC}_{E2} / \Delta\alpha^{TEN}_{E2}$   &  -0.656(19) \\
\multicolumn{2}{l}{{Scalar : Probe field (\B{z})}} &                                                      \\
                                                   & $\Delta\alpha^{SC}_{E3} / \Delta\alpha^{\B{z}}_{E2}$ &  0.471(6) \\
\bottomrule
\end{tabular}

\caption{Observed ratios between the scalar and tensor differential polarizabilities of the two clock transitions in \Yb{} measured at $\lambda=\sm{}$. E2 and E3 refer to the electric quadrupole and octupole clock transitions in \Yb{} respectively. Also included is the ratio of the E3 scalar differential polarizability vs.\ the E2 differential polarizability in a particular magnetic field, \B{z}, used in \cref{sec:absolute_polarizabilities} to establish an absolute value for the differential polarizability. The values shown here for each ratio result from data taken on 3 separate occasions using 2 independent orthogonal sets of fields.  }
\label{tab:ratios}

\end{table}

The ratio of differential scalar to tensor polarizabilities was also measured for both the E2 and E3 clock transitions, $\Delta\alpha^{SC}_{E2} / \Delta\alpha^{TEN}_{E2}$ and $\Delta\alpha^{SC}_{E3} / \Delta\alpha^{TEN}_{E3}$. As above, the frequency shift arising from the differential scalar polarizability was determined by averaging over the frequency shifts recorded in three orthogonal directions.  For the E2 transition, with its large tensor polarizability, the frequency shifts in the orthogonal directions were well separated and it was possible to derive $\Delta\alpha^{SC}_{E2} / \Delta\alpha^{TEN}_{E2}$ directly from these shifts, using \cref{eq:shift}.  Once again, simulations based on observed deviations in the magnitudes of test magnetic fields were used to derive the uncertainties arising from non-orthogonality in the fields of the experiment, as in \ref{appendix:magnetic_field_orthogonality}. These frequency uncertainties, arising from uncertainties in direction, dominated over the statistical uncertainties for the E2 transition. 

For the E3 transition however, the scalar and tensor shifts were both much smaller leading to a higher fractional uncertainty for their statistical noise. To resolve the tensor shift more clearly the frequency shift in \B{y} was measured, corresponding to $\theta=0$ in \cref{eq:shift}, to give an extreme value for the induced tensor shift.  The tensor shift in \B{y} and the scalar shift from the average over three orthogonal directions were therefore used in \cref{eq:shift} to establish $\Delta\alpha^{SC}_{E3} / \Delta\alpha^{TEN}_{E3}$.  

For the tensor shifts, the polarization of the \smh{} beam was also relevant. Observation of the tensor shift magnitude in all the fields (shown in \cref{tab:BFields}) allowed us to determine that the polarization of the \smh{} beam at the ion was aligned along the y axis to within $\pm \SI{1}{\degree}$. This level of uncertainty in the polarisation resulted in additional fractional uncertainties for the E2 and E3 tensor : scalar ratios of \num{5.2e-3} and \num{8.3e-4} respectively: negligible when compared with that due to magnetic field non-orthogonality and quantum projection noise. 
% see https://npluk-my.sharepoint.com/personal/charles_baynham_npl_co_uk/_layouts/OneNote.aspx?id=%2Fpersonal%2Fcharles_baynham_npl_co_uk%2FDocuments%2FOnenote%2F7um&wd=target%28Analysis.one%7C8270892C-A50B-40D6-A9AC-D976F19DF63E%2FE3%20tensor%3Ascalar%2026%5C%2F09%5C%2F16%7CEAFCAC77-B581-404A-A1DD-1B24F120CCAA%2F%29
% and scale this to E2 by using the ratios reported in table 1. 

% section polarizability_ratios (end)
 
\section{Absolute polarizabilities} % (fold)
\label{sec:absolute_polarizabilities}

In order to determine the absolute value of a transition's differential polarizability, knowledge of the strength of electric field from the \smh{} laser at the ion is necessary. Since the electric field $\vec{E}$ from the applied laser was position dependent across the spatial extent of the beam ($S$), the induced shift $\delta\nu(\theta, x,y)$ was measured in a 2D grid of positions, $\vec{x} = (x,y)$, across the beam. 

\begin{figure}[tbh]
	\centering
	\includegraphics[width=\textwidth]{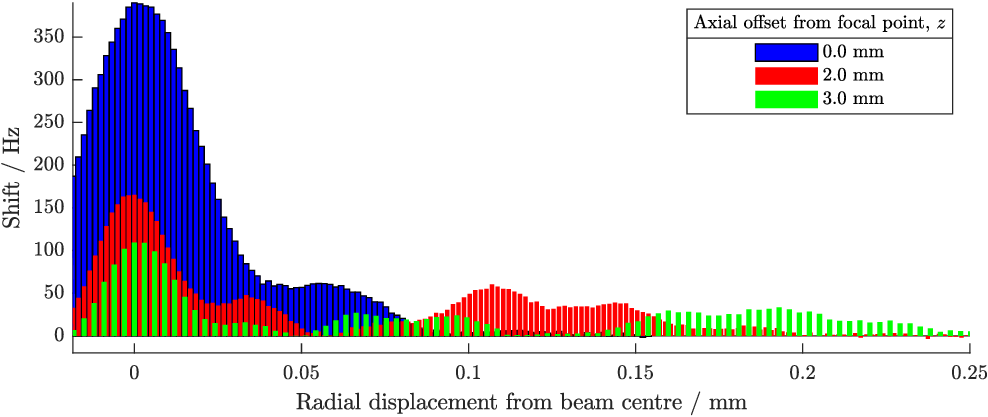}
	\caption{Shift induced by the \smh{} beam for various axial positions of the translation stage. Note that as $z$ nears the focal point of the optical setup \textit{(blue)} the peak shift increases and the power far from beam-centre (visible in these slices as deviations from the Gaussian lineshape) is reduced.}
	\label{fig:focussing}
\end{figure}

Starting from \cref{eq:simple_polar}
\begin{alignat}{2}
	h \, \delta\nu(\vec{x}, \theta) 
		&=\; - \frac{1}{2} \Delta\alpha_{21}(\theta) \; \langle \vec{E}(\vec{x})^2 \rangle 		\nonumber \\
	h \int\delta\nu(\vec{x}, \theta) \, dS	
		&=\; - \frac{1}{2\epsilon_0 c} \Delta\alpha_{21}(\theta) \; \int I(\vec{x}) \, dS
	\, ,
	\label{eq:shift_with_intensity}
\end{alignat}
where $I(\vec{x}) =  \epsilon_0 c \, \langle \vec{E}(\vec{x})^2 \rangle $ is the position-dependent intensity of the \smh{} radiation.
The beam produced by the \smh{} QCL contained high-order components making a Gaussian fit untenable, demonstrated by the 1D slices through the frequency shift profile shown in \cref{fig:focussing}. For this reason, we chose to measure and sum the frequency shifts over the entire extent of the beam. 
By approximating the LHS integral in \cref{eq:shift_with_intensity} by a sum over pixels each of area $A$, and noting that the RHS integral is simply equal to the total power in the beam, $P$, 
\begin{align}
	\label{eq:absolute_polarizability}
	\Delta\alpha_{21}(\theta) 
		&= -  \frac{2\epsilon_0 hc}{P} \sum\limits_{x,y} A \; \delta\nu(\vec{x}, \theta)
	\, .
\end{align}

The polarizability can thus be derived by summing the frequency shifts measured across an entire grid, such as that shown in \cref{fig:colour_map}. Having measured this quantity for one transition in a known quantisation direction $\theta$, the value of any other differential polarizability component can be deduced by means of the ratios given in \cref{sec:polarizability_ratios}. 

The profiles obtained using this method showed good agreement with optical profiles taken of the beam by scanning over a pinhole and measuring a chopped signal with a photo-detector and lock-in amplifier. Unfortunately however, uncertainty of the ion's distance from the focal point prevented a quantitative comparison. 

\subsection{Setup} % (fold)
\label{sub:setup_absolute}

As shown by \cref{eq:absolute_polarizability}, knowledge of four quantities is essential in order to evaluate any component of the ion's differential polarizability: the area of each pixel in a 2D scan, the total power in the beam used to perturb the clock transition, the sum of the resultant shifts across the whole beam, and the ratio linking the polarizability component in the measured direction to the direction of interest. The following sections will detail how each of those quantities was controlled and measured. 

\subsubsection[Sum of frequency shifts]{Sum of frequency shifts over laser profile : $\sum\limits_{x,y} \delta\nu(\vec{x}, \theta)$} % (fold)
\label{ssub:sum_of_shifts}

\begin{figure}[tbh]
	\centering
	\includegraphics[width=\textwidth]{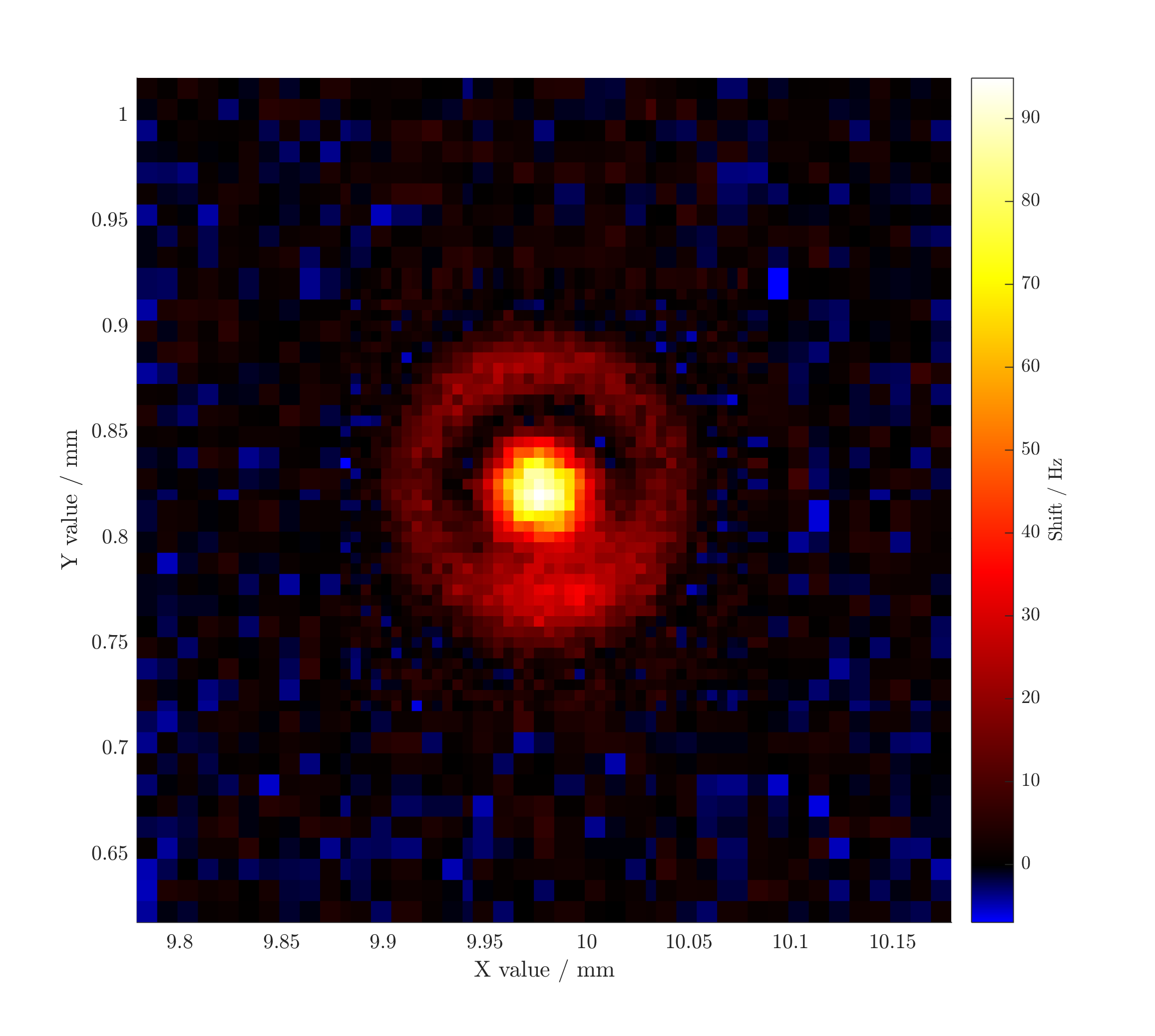}
	\caption[Colour map of the induced shift]{Colour map of the induced frequency shift at a large range of $(x,y)$ positions. Each data point represents the average differential shift observed between \SI{30}{\second} interleaved locks to both the shifted and unshifted quadrupole clock transition in \ch{Yb+} with the ion's quantization axis in the $\hat{z}$ direction and the electric field in the $\hat{y}$ direction. The scan was performed in a spiral manner outwards and then inwards. }
	\label{fig:colour_map}
\end{figure}

The spatial profile of the beam was mapped out by translating the \smh{} beam across the ion and recording the frequency shifts induced at each $(x,y)$ position, as shown in \cref{fig:colour_map}. This was done using the E2 transition since its polarizability is $\sim10\times$ greater than that of the E3 transition, allowing the shift to be resolved more quickly. The large number of points required to cover the entire transverse beam profile meant that each scan could take around two days. By operating in a magnetic field such that $\theta = \pi/2$ (field \B{z} in \cref{fig:experimental_setup}) we minimised our sensitivity to changes in the background field affecting the tensor shift during this time.%

\cref{fig:colour_map} shows the shift induced at a range of positions of the translation stage for fixed z-position, covering a large area across the beam. 

\paragraph{Summing uncertainty} % (fold)
                                 
In order to approximate the integral over all $(x,y)$ points in the beam, we used the simple method of summing all the measured shifts in a 2D grid, weighted by their corresponding grid square areas (described by \cref{eq:absolute_polarizability}), since the inner region of the grid was scanned using smaller position steps than the outer. 

Simulations of the error in the summed shift deduced by this simple method were performed, assuming a beam profile with either a Gaussian shape or an analytical function that closely approximated the observed profile. The results in both cases indicate that the penalty incurred by using the sum of pixelated shifts instead of the complete, smooth beam integral is surprisingly small even for large (i.e.~approaching the beam-waist) grid spacings and also for off-centre grids. 
The contribution of the grid summing method to our total uncertainty can therefore be neglected at the $10^{-4}$ level.

\paragraph{Statistical} % (fold)

% See Onenote page: onenote:https://npluk-my.sharepoint.com/personal/charles_baynham_npl_co_uk/Documents/YbWeirdness/7um%20laser.one#E2%20and%20E3%20absolute%20polarizability%2011/11/16&section-id={D7910ED3-8D70-483D-9F0E-26C24B2AA077}&page-id={A333410B-EA17-4830-9272-173E9556A40C}&end

Each point in \cref{fig:colour_map} was typically scanned for \SI{30}{\second}.  At this time-scale, the quantum projection noise of the interleaved E2 clock transition is around \SI{3}{\hertz}. Including this per-point uncertainty results in a fractional statistical uncertainty of \SI{0.8}{\percent} on the sum of shifts over the whole grid. 

\paragraph{Beam capture}
\label{para:beam_capture}

\begin{figure}[tb]
	\centering
	\includegraphics[width=\textwidth]{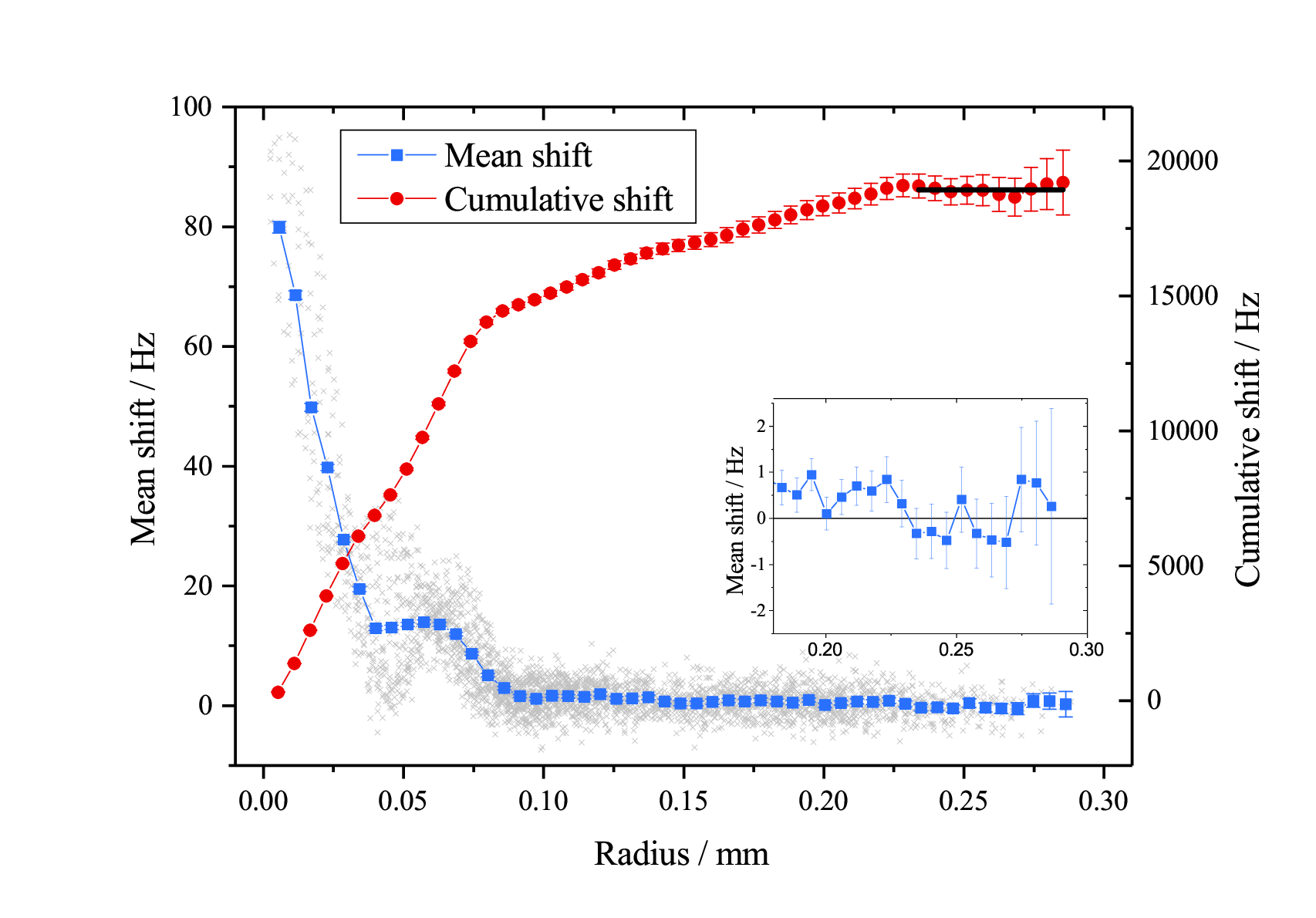}
	\caption{\emph{(blue)} Mean shift of data in \cref{fig:colour_map} for circular bins spaced by $\delta r = \SI{6}{\um}$. The gray crosses represent the individual data points that contributed to this mean. 
	The inset shows a close-up view of the measurements at the largest radii, consistent with no power in the \smh{} beam being present. 
	\emph{(red)} Cumulative sum of the binned shifts, with black line showing the weighted mean of the last few points. }
	\label{fig:beam_capture}
\end{figure}

The method of considering the sum of shifts in
\cref{eq:absolute_polarizability} relies on probing the beam over its entire
extent; power missed by the 2D grid would cause an underestimate of the
differential polarizabilities. To judge how well all the frequency shifts have
been captured, \cref{fig:beam_capture} shows the shifts and their mean, as a
function of distance from the beam centre.  The plot is derived from the data
in \cref{fig:colour_map} and the standard error on the mean shifts comes from the
\SI{3}{\hertz} statistical noise on each of the points in the average
mentioned above.  At large radii, the standard errors are larger simply
because there are fewer points contributing to the mean.  The shifts at large
radii are consistent with zero and we assume that there is no further power in
the beam beyond the maximum radius measured.

This does not necessarily mean that our experiment captured all the shifts,
however, as we took data in a square grid that only went out to the highest
radii in the corners of the grid.  We therefore need to estimate the total sum
of shifts that would have been measured if we had sampled out to the highest
radii across the full circular area.  We do this by treating the circular area
as a series of annular rings around the beam centre, each of which is assigned the mean shift at that radius around the whole ring. 
Note that this does not assume that the
beam itself is circularly symmetric, only that the data contributing to the
mean shift at each radius was sampled evenly around that ring. 

The cumulative sum of shifts is plotted as a function of radius in
\cref{fig:beam_capture} and is seen to level off beyond a radius of
\SI{220}{\um}.  The cumulative sum of shifts is then estimated
from a weighted fit over the last few points in the plot to be
\SI{18931(43)}{\hertz} across the whole circular area.  This number has been
scaled by area to correspond to a grid-sum over \SI{5}{\um} pixels so that it can be
readily compared to the actual sum of shifts, \SI{18712}{\hertz}, measured
over the square grid in \cref{fig:colour_map}. This implies that a correction
factor of
\newcommand{\missedPowerCorrectionFactorNum}{1.012}%
\newcommand{\missedPowerCorrectionFactorFracError}{2}% x 10^-3
\newcommand{\missedPowerCorrectionFactor}{\num{1.012(2)}}%
\missedPowerCorrectionFactor{} is required in our cumulative sum to account for
shifts occurring beyond the square area sampled by our experiment. 

% subsubsection sum_of_shifts (end)

\subsubsection[Area]{Area : $A$} % (fold)
\label{ssub:Area}

% X data
% ======
% Onenote link: onenote:https://npluk-my.sharepoint.com/personal/charles_baynham_npl_co_uk/Documents/YbWeirdness/7um%20laser.one#09-01-2016%20-%20Nanomover%20x%20step%20analysis&section-id={D7910ED3-8D70-483D-9F0E-26C24B2AA077}&page-id={C53187ED-302E-49D9-9E33-9FAD0467D55B}&end
% XYStepping 06-01-17-001-NanomoverX_tentOn.txt
% Using this value for w, a nominal 1 mm step would really be 0.994709 mm
% Code version: 17-Aug-2017 16:40:55 - #697563 - plotFringes.m
% Y Data
% ======
% Onenote link: onenote:https://npluk-my.sharepoint.com/personal/charles_baynham_npl_co_uk/Documents/YbWeirdness/7um%20laser.one#06-01-2017%20-%20Nanomover%20y%20step%20matlab%20analysis&section-id={D7910ED3-8D70-483D-9F0E-26C24B2AA077}&page-id={E76AA676-F5CA-41FC-8F26-86C55CBA758E}&end
% XYStepping 05-01-17-001-NanomoverY_tentOn.txt
% Using this value for w, a nominal 1 mm step would really be 0.978401 mm
% Code version: 17-Aug-2017 16:39:32 - #697563 - plotFringes.m

To determine the area, $A$, of the grid squares used in \cref{eq:absolute_polarizability}, a Michelson interferometer setup with $\lambda = \SI{935}{\nano\meter}$ was used to monitor the step size of the NanoMover translation stage and its repeatability. 

The fringes produced by changes in one arm of the interferometer due to movement of the translation stage were measured by a photodiode and analog--digital converter to be analysed digitally. 

\begin{figure}[tb]
    \centering
    \begin{subfigure}[t]{0.5\textwidth}
        \centering
        \includegraphics[height=7cm]{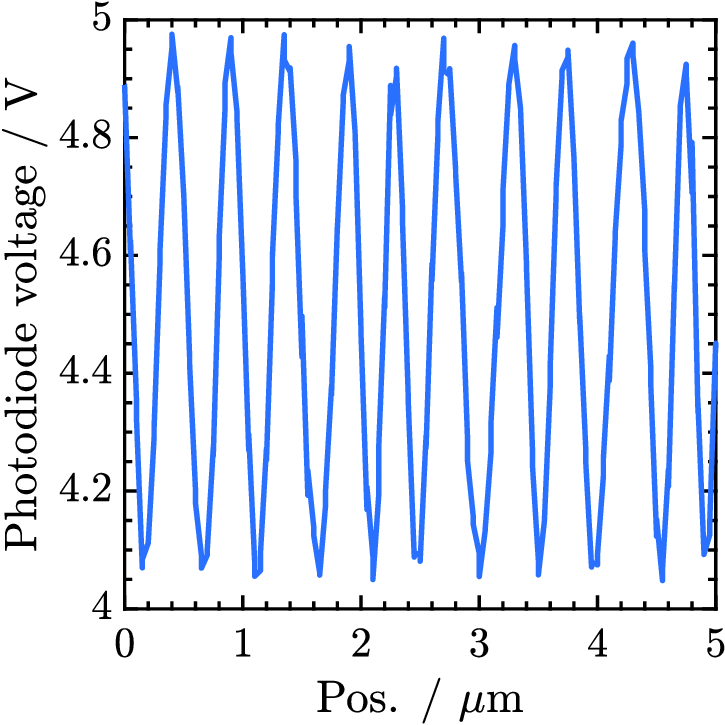}
    \end{subfigure}%
    ~ 
    \begin{subfigure}[t]{0.5\textwidth}
        \centering
        \includegraphics[height=7cm]{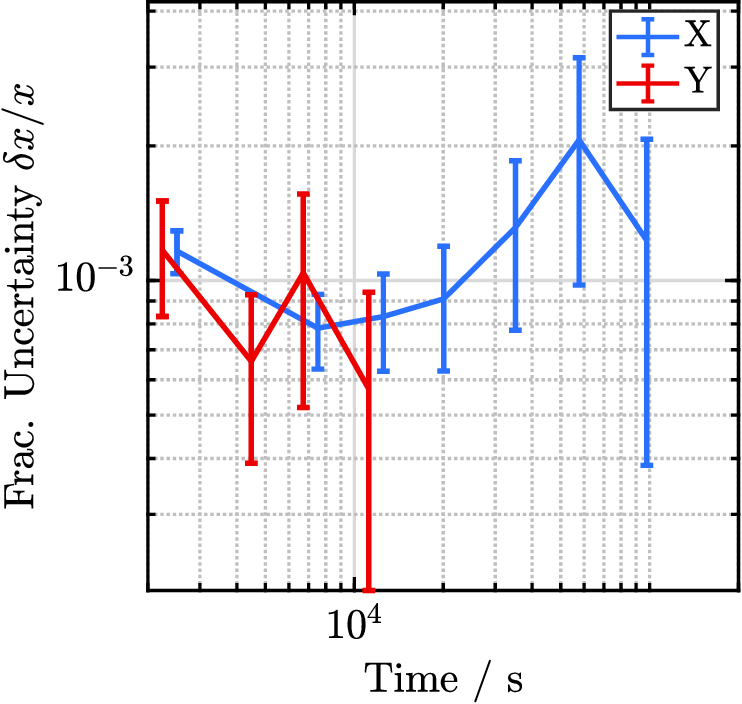}
    \end{subfigure}
    \caption[Allan deviation of fractional uncertainty in translation-stage step size]
		{
		\textit{(left)} Typical data for a single run of a Michelson interferometer setup with \SI{935}{\nano\meter} light, for translation-stage movement in the X or Y directions. Each run took \SI{2200}{\second}: \SI{58}{\hour} of data were taken for the X direction and \SI{7.5}{\hour} for the Y.
		\textit{(right)} Allan deviation of fractional uncertainty in translation-stage step size over time.
		}
	\label{fig:fringes}
\end{figure}

We ran datasets in the same conditions as when probing the ion with the interferometer set up for monitoring either $x$ or $y$ step sizes: see \cref{fig:fringes}. 
The fractional instability in step sizes produced by the NanoMover was below  $2\times10^{-3}$  over more than 8 hours: the time-scale of the \smh{} laser scans. We found that nominal \SI{10}{\micro\meter} steps in the $x$ and $y$ directions, as programmed in the control software, produced $\SI{9.95(2)}{\micro\meter}$ and $\SI{9.78(2)}{\micro\meter}$ steps respectively. 

% subsubsection step_size (end)

\subsubsection[Beam power]{Beam power : $P$} % (fold)
\label{ssub:beam_power}

In order to determine the beam power at the ion, knowledge of two elements is necessary: firstly, its power before entering the trap and, secondly, the attenuation caused by the windows of the vacuum chamber. In addition, an understanding of the stability of the power over the time between the polarizability scans and the power measurement is also required. 

\paragraph{Traceability}

A pick-off mirror in the \smh{} beam path enabled continuous monitoring of laser power stability and a NIST-calibrated optical power transfer standard could be inserted into the main beam to measure its power with high accuracy. As optical power instrumentation suffers errors introduced by spectral non-uniformity, accurate measurement of laser power typically requires instrumentation calibrated against a primary standard at the wavelength to be studied. Due to power non-linearities, it is also necessary to calibrate near the power to be measured, typically within one-two decades depending on instrument performance. 
Direct calibrations of laser power instrumentation at the \SI{20}{\milli\watt} level at \sm{} are not presently available from any calibration lab and, while spectrometer ``bridging'' measurements of the device or its coating are possible, this would introduce an unacceptably high uncertainty for the present application.

As such, a device coated with a Vertically Aligned Nanotube Array (VANTA) developed at NIST was used as the optical power standard. VANTA coatings show superior spectral uniformity in the VIS-FIR regime\cite{Theocharous2013} and thereby permit a calibration performed against a primary laboratory standard at an arbitrary wavelength to serve as a standard at another arbitrary wavelength, with corrections for the small spectral variation and uncertainty contribution of the VANTA coating measurement.

Use of this device allowed the power in the \smh{} beam to be measured as \SI{17.52(16)}{\milli\watt}. The uncertainty here refers to traceability to primary standards: the short-timescale instability of the device is much lower. 

\paragraph{Window attenuation}

% See December 2016, 22-12-16 in Onenote

To know the power in the plane of the ion, it was necessary to measure the attenuation experienced by the \smh{} beam as it passed through the \ch{MgF2} window of the vacuum chamber.

Since access to the in-vacuum side of the trap window was impractical, an identically manufactured window from the same batch as that mounted to the chamber was used to characterise absorption. The \smh{} beam was aligned such that it passed through the test window at the same angle and with the same beam waist as in the main experiment. The attenuation through the window was measured using an IR power meter. This test was then repeated for different beam positions on the window.

The attenuation caused by the window was found to be insensitive to position at the \SI{0.2}{\percent} level. Taking this into account, the transmission of the \ch{MgF2} window at \sm{} was found to be 0.5101(14).

\paragraph{Stability}

Although we stated the surprisingly impressive fractional power stability of the \smh{} beam of \num{6E-5} at timescales of \SIrange{1000}{10000}{\s}, this was not the number used here because it is not the most relevant timescale for the absolute polarizability experiment. Each grid scan took around 30 hours, and 2 separate grids were taken. To combine the data and treat it as a single dataset (as shown in \cref{fig:colour_map}), we needed to consider any power fluctuations over the total time of measuring and until the calibrated power meter could be inserted at the ion's position. 

The monitored power showed fluctuations at a level of \SI{0.5}{\percent} over several days: to be conservative, we therefore used \SI{1}{\percent} for the power fluctuations in our uncertainty analysis of the power at the ion. 

% subsubsection beam_power (end)

\subsubsection[Ratio to probe]{ Ratio : $ \Delta\alpha^{SC} / \Delta\alpha^{\B{z}}_{E2} $ } % (fold)
\label{ssub:ratio_to_probe_uncertainty}

The previous sections have described the measurements undertaken to derive the absolute differential polarizability in the field \B{z} where $\theta = \pi/2$. To link this measurement to the scalar and tensor components of the two transitions requires knowledge of the ratio between these components and the differential polarizability in \B{z}. 

The determination of these ratios was subject to the same considerations as those required when evaluating the ratios presented in \cref{sec:polarizability_ratios}. For the scalar components of the E2 / E3 transition, the fractional uncertainty contribution was \SI{2.0}{\percent} / \SI{1.3}{\percent}. In order to derive the absolute values of the tensor shifts of the two transitions, the scalar : tensor ratio measurements previously presented in \cref{tab:ratios} were used. 

% subsubsection ratio_to_probe_uncertainty (end)

\subsection{Results} % (fold)
\label{ssub:absolute_results}

The results of the measurements described in \cref{sec:absolute_polarizabilities} are shown in \cref{tab:uncertainty_budget}. These data, in combination with \cref{eq:absolute_polarizability}, are used to calculate the absolute values of the scalar and tensor components of the differential polarizabilities of the two transitions. 

\begin{table}[ht]
%
% All the numbers for this table come from the Excel worksheet stored in this repo. 
% The version used to generate this table was #f86ef1fd41
%	
	\makebox[\textwidth][c]{%	
		\begin{tabular}{@{}l@{\hspace{5mm}}rrr@{}}
		\toprule
Contribution                                                         & Value  & \multicolumn{2}{r}{Fractional uncertainty ($10^{-3}$)} \\
\midrule
Area of pixels / $\si{\micro\meter}^2$                                        & 24.3                              &                     & 3                                       \\
Power at ion / $\si{\milli\watt}$                                             & 8.94                              &                     & 14                                      \\
Sum of shifts / $\si{\hertz}$                                                 & -18712                            &                     & 8                                       \\
Correction to sum of shifts                                                   & \missedPowerCorrectionFactorNum{} &                     & \missedPowerCorrectionFactorFracError{} \\
$\Delta\alpha^{SC}_{E3} / \Delta\alpha^{\B{z}}_{E2}$                          & 0.471                             &                     & 13                                      \\
$\Delta\alpha^{SC}_{E2} / \Delta\alpha^{SC}_{E3}$ \textit{(used for E2 only)} & 9.13                              &                     & 15                                      \\ \midrule % \addlinespace[0.5em] % Vertical gap
                                                                              &                                   & \textit{Quadrupole} & \textit{Octupole}                       \\ %\cmidrule(lr){3-4}
\textbf{Differential polarizability @ \sm{} }                                 & scalar                            & 7.79(20)            & 0.854(18)                               \\
$\quad / \: \si{\joule\metre\squared\per\volt\squared} \times 10^{-40}$       & tensor                            & -11.88(46)          & -0.2067(67)                             \\ \bottomrule
		\end{tabular}
	}

	\caption{	Uncertainty budget and result of the determination of the absolute differential polarizabilities at \sm{} of the two clock transitions. 
				Tensor components are derived from the scalar components using the ratios presented in \cref{tab:ratios}.}
	\label{tab:uncertainty_budget}

\end{table}

% \begin{figure}[tbh!]
% 	\centering
% 	% \missingfigure{Numerical integration over Gaussian profile beam simulation}
% 	\caption{Simulated error in numerical integration over a Gaussian profile for varying grid spacing}
% 	\label{fig:grid_spacing}
% \end{figure}

% subsection combined_results (end)
% subsection absolute_polarizabilities (end)

\section{Estimating BBR frequency shift} % (fold)
\label{sec:deducing_bbr_frequency_shift}

% The analysis used to produce these numbers can be reproduced by running Analysis/CFAB/RunFinalAnalysis.m
% The version used to get these values for the E3 was #043325d1.
% These results are summarised in the 2016 Labbook Onenote, under
%	onenote:https://npluk-my.sharepoint.com/personal/charles_baynham_npl_co_uk/Documents/YbWeirdness/7um%20laser.one#E2%20with%20NIST%20corr  & section-id={D7910ED3-8D70-483D-9F0E-26C24B2AA077} & page-id={40A206F3-AD64-4F30-BA15-A8D2736DAD4C} & end
% and
% 	onenote:https://npluk-my.sharepoint.com/personal/charles_baynham_npl_co_uk/Documents/YbWeirdness/7um%20laser.one#E3%20with%20NIST%20corr & section-id={D7910ED3-8D70-483D-9F0E-26C24B2AA077} & page-id={24F42D93-E5D2-4E47-B669-57FCF1B0F52C} & end
%
\newcommand{\quadBBRShift}{-338} % mHz
\newcommand{\quadBBRShiftU}{3.5} % percent
\newcommand{\octBBRShift}{-43.4} % mHz
\newcommand{\octBBRShiftU}{2.1} % percent
\newcommand{\quadDCPol}{5.89(30)} % 10^-40 JM^2/V^2
\newcommand{\octDCPol}{0.859(18)} % 10^-40 JM^2/V^2
\newcommand{\quadDynCorr}{+0.135(22)} % unitless
\newcommand{\octDynCorr}{-0.00245(16)} % unitless

The differential scalar polarizability at \sm{} allows the atomic frequency shift to be deduced close to the peak of the room-temperature black-body radiation spectrum. 
The radiation spectrum, however, extends to both higher and lower wavelengths, and so the total frequency shift from a black body at room temperature also depends on the polarizability's wavelength dependence.

A basic theoretical model can be used to estimate the scalar polarizability $\alpha$ of state $\ket{\gamma J}$ at different wavelengths by summing over the oscillator strengths $f$ of all the electric dipole transitions connecting state $\ket{\gamma J}$ to states $\ket{\gamma' J'}$ using~\cite{Sobelman2012}:
\begin{equation}
	\label{eq:theory_levels}
	\alpha(\gamma J) = \frac{e^2}{m_e} \sum_{\gamma'J'} \left( \frac{f(\gamma J; \gamma' J')}{\omega^2_{\gamma J:\gamma' J'} - \omega_L^2}\right)
\end{equation}
where $\omega_{\gamma J:\gamma' J'}$  is the atomic transition frequency and $\omega_L$ is the frequency of the applied electric field. $e$ and $m_e$ are the electronic charge and mass.  The oscillator strengths are taken from~\cite{Pinnington1997,Lea2006,Hayes2009,Taylor1997,Biemont1998}, except for the leading term in the summation for each state. 

The oscillator strengths of the leading terms for the excited and ground states are allowed to vary around their theoretical values%
\footnote{The range of these variations was chosen such that the weights assigned later were non-negligible or, for transitions with well-known, experimentally measured oscillator strengths, to cover a 3-sigma range.}
and, for each combination of oscillator strengths, the differential polarizability at all wavelengths is calculated giving curves such as those shown in \cref{fig:polarizability-and-BBR}. For the E2 transition, the line-strengths of transitions at wavelength $\lambda = $ \SIlist{329; 2438}{\nano\meter} were chosen to be varied. For the E3, the \SIlist{329;265;3432}{\nano\meter} transitions' line-strengths were used. 
The \SIlist{329;265}{\nano\meter} transitions were chosen because their line-strengths have the largest contributions to the polarizability of the lower clock state, \ch{^2S_{1/2}}, at \sm{}. The \SIlist{2438;3432}{\nano\meter} line-strengths have the largest contributions to the upper clock state polarizabilities for the E2 and E3 transitions respectively. The UV transitions are so far-detuned from \sm{} that their contribution to the differential polarizability is a near-constant offset at IR wavelengths. The transitions closer to IR however have a much larger effect on the shape of the differential polarizability in the BBR region.  

The validity of each combination of oscillator strengths is judged by comparing the predicted polarizability at \sm{} to our experimental observations for each clock state. Each polarizability curve is thus assigned a weight, calculated from its probability of lying in a normal distribution centred at our measurement with our experimental uncertainty. A weighted mean of all these curves gives our predicted differential scalar polarizability as a function of wavelength, with its uncertainty their standard deviation. 

% \todo{Some sort of figure would probably be helpful here – preferably with the polarizabilities matching our exp observations, rather than the raw Biemont polarizabilities plotted here.}

The BBR frequency shift can then be estimated by integrating, across all wavelengths, the product of the predicted differential scalar polarizability and the electric field squared.  We derive the frequency shift from a black body at \SI{298}{\kelvin} for the E2 transition to be \SI{\quadBBRShift}{\milli\hertz} with a fractional uncertainty of \SI{\quadBBRShiftU}{\percent}, and for the E3 transition to be \SI{\octBBRShift}{\milli\hertz} with a fractional uncertainty of \SI{\octBBRShiftU}{\percent}.  For comparison with other work, the model also predicts the dc scalar differential polarizabilities  (as shown in \cref{tab:final_results}) to be 
$\SI{\quadDCPol E-40}{\joule\metre\squared\per\volt\squared}$ for the E2 transition and 
$\SI{\octDCPol  E-40}{\joule\metre\squared\per\volt\squared}$ for the E3 transition, and the dynamic corrections (using the definition from~\cite{Porsev2006}) at \SI{298}{\kelvin} to be \quadDynCorr{} and \octDynCorr{} for the E2 and E3 transitions respectively. 

The oscillator strength combinations that best predict our experimental
polarizabilities cannot be used to draw conclusions about the true oscillator
strengths of transitions in the ion however. In the dc limit, the denominator
in \cref{eq:theory_levels} becomes constant so the contribution from each
transition is a constant value. Varying any of these would simply lead to an
increase or decrease in the value of the dc polarizability, and so it would
not matter which you chose. This is approximately true for $\lambda = \sm$
when considering the UV transitions in \Yb{}, and so our analysis does not
separately predict oscillator strengths of the individual transitions.

% section deducing_bbr_frequency_shift (end)

\section{Conclusion} % (fold)
\label{sec:conclusion}

We have directly measured the differential polarizabilities for the E2 and E3 clock transitions in \Yb{} using a perturbing field at $\lambda = \SI{7.17}{\micro\metre}$, a wavelength in the region typical of room temperature BBR spectra. 

We presented ratios between the tensor and scalar components of the differential polarizabilities of both transitions and fully characterized the \smh{} beam's intensity profile, allowing the absolute values of the differential polarizabilities to be deduced, all at the few percent level of uncertainty. 
A basic theoretical model was used to predict the polarizabilities at other wavelengths, thus allowing the BBR shifts to be calculated. 

% From PRL 94, 230801 (2005)
% The quadrupole moment of the 5d 2D3=2 state is measured
% to be 9.32(48)E-40 C m2 and from the quadratic Stark shift the relevant scalar and tensor polarizabilities
% are determined to be E2 SEC: -6.9(1.4)E-40 J m2=V2 and E2 TEN: -13.6(2.2)E-40 J m2=V2, respectively. 

\begin{table}[t]

	% Display uncertanties as 1.5 +- 0.2 instead of 1.5(2)
	% Note that latex's grouping rules means that this applies in this table only
	\sisetup{separate-uncertainty=true}

	\centering

	\begin{tabular}{lrl@{~}ll@{~}l@{}}
	\toprule
              &                 & \multicolumn{2}{l}{\textit{Quadrupole}}                                                                     &  \multicolumn{2}{l}{\textit{Octupole}} \\
              & \phantom{12345} & \multicolumn{4}{c}{$\Delta\alpha^{SC}_{DC}$/ $\si{\joule\metre\squared\per\volt\squared} \times 10^{-40} $} \\ \cmidrule(lr){3-6}
This work     &                 & \num{\quadDCPol}                                                                                            &                                        &  \num{\octDCPol} &                      \\
Previous work &                 & \num{6.90(140)}                                                                                             &  \cite{Schneider2005}                  &  \num{0.888(16)} & \cite{Huntemann2016} \\
\bottomrule
%\textbf{Total}&& 5.89(29)& 0.872(12)\\
%\textit{Frac. clock frequency uncertainty}
%             & $\mathit{10^{-17}}$ & $\mathit{10^{-18}}$
\end{tabular}

\caption{Values for the dc scalar differential polarizabilities %
% $\alpha^{SC}_{DC}(excited) -\alpha^{SC}_{DC}(ground)$
of the E2 and E3 transitions extrapolated using the method described in \cref{sec:deducing_bbr_frequency_shift}, shown for comparison with previous work. 
}
\label{tab:final_results}

\end{table}

Further work with more sophisticated theoretical models may improve upon the values presented here.  Nevertheless, our basic model already reveals that the polarizability measurements at \sm{} greatly reduce the uncertainty of the BBR frequency shift for the E2 transition from approximately \SI{20}{\percent} \cite{Schneider2005} to \SI{\quadBBRShiftU}{\percent}.  This is significant as the BBR shift dominates the uncertainty budget for the E2 transition frequency in \Yb{}.
We note that our predicted dc differential scalar polarizability for the E2 transition is significantly different from a recent \textit{ab initio} theoretical calculation \cite{Roy2017} that predicts 
$\Delta\alpha^{SC}_{DC} = \SI{7.8(5) E-40}{\joule\meter\squared\per\volt\squared}$. % DOI: 	10.1088/1361-6455/aa8bae
Nevertheless, we note reasonable agreement with totally independent experimental results for the dc differential scalar polarizabilities for both the E2 and E3 transitions, as shown in \cref{tab:final_results}. 
 
These new measurements show that limits on the fractional frequency uncertainty caused by room-temperature black-body radiation can be reduced to low parts in $10^{17}$ on the E2 transition and low parts in $10^{18}$ for the E3 transition. These uncertainties assume a temperature uncertainty in the ion's environment of \SI{1}{\kelvin} or better, commonly achieved experimentally~\cite{Nisbet-Jones2016,Dolezal2015}. 

% section conclusion (end)

\section*{Acknowledgements}

We acknowledge financial support from the UK Department for Business, Energy and Industrial Strategy as part of the National Measurement System programme; the European Metrology Research Programme (EMRP) project SIB04-Ion Clock; and the European Metrology Programme for Innovation and Research (EMPIR) project 15SIB03-OC18. This project has received funding from the EMPIR programme co-financed by the Participating States and from the European Union’s Horizon 2020 research and innovation programme.  The EMRP is jointly funded by the EMRP participating countries within EURAMET and the European Union.

\appendix
\renewcommand{\thesubsection}{\Alph{section}.\arabic{subsection}}

\section{Magnetic field orthogonality}
\label{appendix:magnetic_field_orthogonality}

\setcounter{section}{1}

In order to determine the ratio between scalar and tensor differential
polarizabilities of the E2 and E3 transitions, knowledge of the orientation of
the magnetic field experienced by the ion was necessary. \Cref{eq:shift} shows
that, for a given set of quantum numbers, a state's tensor shift is
proportional to $(3\cos^2\theta - 1)$ where $\theta$ is the angle between the
ion's quantization axis (set by the magnetic field direction) and the
perturbing \smh{} laser's electric field. To obtain the scalar components of
the differential polarizability, we taken advantage of the fact that this term
sums to zero over three orthogonal magnetic field directions and so the mean
of the shift to the clock transition's frequency in these three fields is
entirely due to the scalar component of the differential polarizability.

In this appendix we will present the method used both to measure the fields and to characterise the uncertainty introduced. This method takes two parts: in subsection~\ref{sub:determination_of_magnetic_fields} we summarise the method used to measure and control the magnetic field at the ion's position and in subsection~\ref{sub:characterization_of_uncertainty} we evaluate the resultant impact of error in these field directions on the shifts caused by incident \smh{} radiation. 
We will find that this contribution is the dominant cause of uncertainty for the shifts to the E2 transition due to its large tensor differential polarizability, whereas the E3's smaller tensor component means that statistical uncertainty dominated for these shifts. 

Although the analysis is described with one set of orthogonal fields, the polarizability ratios shown in \cref{tab:ratios} are an average of measurements taken in two independent sets of three field orientations as a consistency check. Labelling the magnetic fields in these two sets as \B{A1}, \B{A2}, \B{A3} and \B{B1}, \B{B2}, \B{B3}, the total set of fields used for this analysis is given by:
%
% Fields used were B789, B123, -By and Bz in lab notation. 
% A1,2,3 are B789
% B1,2,3 are B123
% -By becomes Bz
% Bz becomes By

% Here are the fields from RMG's matlab:
% B =       [ B7, B8, B9, BX, BY, BZ, B1, B2, B3];
% B_ideal = [ -8.3709,-5.2947, 0.9347, 10,0,0,-9.3587,-2.4515,-2.5528;
%             -2.3561, 5.2985, 8.1856, 0,10,0,-3.2700, 3.6846, 8.7173;
%              4.9373,-6.6251, 5.6677, 0,0,10, 1.3121,-8.9674, 4.1824];
\begin{table}[h]
	
	\centering

	\begin{tabular}{llllllllll}
	\toprule
% *******************
% 29-Mar-2018 17:02:49 - #Response_sent_to_authors-27-g8e1a94 - outputFields.m
% *******************
                                         & \B{A1} & \B{A2} & \B{A3} & \B{B1} & \B{B2} & \B{B3} & \B{x} & \B{y} & \B{z} \\ \midrule
\textit{x} component / \si{\micro\tesla} & -8.37  & -5.29  & 0.93   & -9.36  & -2.45  & -2.55  & 10.00 & 0.00  & 0.00  \\
\textit{y} component / \si{\micro\tesla} & 4.94   & -6.63  & 5.67   & 1.31   & -8.97  & 4.18   & 0.00  & 10.00 & 0.00  \\
\textit{z} component / \si{\micro\tesla} & 2.36   & -5.30  & -8.19  & 3.27   & -3.68  & -8.72  & 0.00  & 0.00  & 10.00 \\
% *******************
% Probe beam propegation:
%     0.5299
%          0
%    -0.8480

% *******************
	\bottomrule
	\end{tabular}
	\caption{Magnetic fields used in this work. All fields have a magnitude of \SI{10}{\micro\tesla} and are given w.r.t.\ the axes shown in \cref{fig:experimental_setup}.}
	\label{tab:BFields}
\end{table}

\subsection{Determination of magnetic fields} % (fold)
\label{sub:determination_of_magnetic_fields}

To control the fields at the ion, three orthogonal pairs of coils in the Helmholtz configuration are fitted to the ion trap (see \cite{Nisbet-Jones2016} for a full description).  Current can be passed through these coils to create arbitrary fields at the ion's position.
To characterise these fields, the ion itself is used as a sensor, with
Zeeman splittings of the magnetically sensitive $\ch{{}^2S_{1/2} \; (F = 0)} \rightarrow \ch{{}^2D_{3/2} \; (F = 2)}, \Delta m_F = \pm1, \pm2$ transitions revealing the magnitude of applied magnetic fields. As the ion's Zeeman splittings do not give
information about field direction, it is necessary to test field magnitudes over a
wide range of applied magnetic field angles: we used three independent sets
of three orthogonal field directions.

The field experienced by the ion is modelled by the following:
\begin{align}
	\label{eq:BFieldModel}
	\vec{B}(\vec{I}) &= \mat{J} \vec{I} + \vec{B}[0] \nonumber\\
			&\equiv
\begin{bmatrix}
    \frac{\partial \B{x}}{\partial \vec[1]{I}}  & \frac{\partial \B{x}}{\partial \vec[2]{I}} & \frac{\partial \B{x}}{\partial \vec[3]{I}} \\
    \frac{\partial \B{y}}{\partial \vec[1]{I}}  & \frac{\partial \B{y}}{\partial \vec[2]{I}} & \frac{\partial \B{y}}{\partial \vec[3]{I}} \\
    \frac{\partial \B{z}}{\partial \vec[1]{I}}  & \frac{\partial \B{z}}{\partial \vec[2]{I}} & \frac{\partial \B{z}}{\partial \vec[3]{I}}
\end{bmatrix}
\begin{bmatrix}
    \vec[1]{I} \\
    \vec[2]{I} \\
    \vec[3]{I}
\end{bmatrix}
+
\begin{bmatrix}
     \vec[x]{B}[0] \\
     \vec[y]{B}[0] \\
     \vec[z]{B}[0]
\end{bmatrix}
\end{align}
where $\vec{B}[0]$ is the background magnetic field and $\vec{I}$ is a vector of currents (although we hope that current coil 1 produces exclusively field in direction $\unit{x}$, this is not assumed: the subscripts are therefore distinct). The Jacobian matrix element $J_{i,j}$ quantifies the effect of current in coil $j$ on the magnetic field in direction $\unit{i}$. 

The values of the 12 coefficients here are found using a non-linear least-squares algorithm to minimise the residual
\begin{equation}
	\label{eq:BFieldResidual}
	\sum_\mu R_\mu^2 = \sum_\mu  ( J \vec{I}[\mu] + \vec{B}[0] )^2  - B_\mu ^2
	\quad.
\end{equation}
where $B_\mu$ is the magnitude of the field measured for the set of coil currents $\vec{I}[\mu]$. 
Since the Jacobian matrix $\mat{J}$ is related to the trap structure, we find that these do not significantly vary over time. A one-off measurement in 51 different field magnitudes and directions was therefore performed to produce values of these coefficients, and thereafter only the value of the magnetic field offset $\vec{B}[0]$ is allowed to change in the fit. 

Before each polarizability measurement described in
\cref{sec:polarizability_ratios,sec:absolute_polarizabilities} the background
field is determined by measuring the actual magnitude of the magnetic field
caused by the 3 different sets of 3 coil currents intended to produce the orthogonal
fields listed in \cref{tab:BFields}. This set of 9 measurements is used to constrain the 3 free parameters
of $\vec{B}[0]$ and calculate the actual currents required to produce our
desired sets of fields. The validity of this calculation is then confirmed by
re-probing the Zeeman components using this corrected set of currents, before the 
polarizability measurement is performed. 
After each polarizability measurement
(typically lasting several hours) this magnetic field measurement is repeated
and the change in field strength over the course of the polarizability measurement is
determined. To be conservative, we take the maximum deviation of the field
strength of any of the 9 fields from their initial value to be
the uncertainty ($\sigma_B$) on each of the 3 magnetic fields used in the differential
polarizability measurements.

% subsection determination_of_magnetic_fields (end)

\subsection{Characterization of uncertainty} % (fold)
\label{sub:characterization_of_uncertainty}

% **** Non-orthogonality ****
% 0.250uT Tolerences, B=10.00uT, n=10 Boff = +/-0.60, Bmult = 1.00%
% Centre = 0.30, width = 1.6423-Mar-2018 14:40:58 - #m1.0-112-gc57030-D - ResidualTensorShift.m
% B field uncertainty: 0.250 uT
% Sum of offset + 1-sigma = 1.940 Hz
% **************************
\newcommand{\nonOrthogOffset}{\SI{0.30}{\hertz}}
\newcommand{\nonOrthogRange}{\SI{1.64}{\hertz}}
\newcommand{\nonOrthogBField}{\SI{10}{\micro\tesla}}
\newcommand{\nonOrthogBFieldUncert}{\SI{0.250}{\micro\tesla}}
\newcommand{\nonOrthogTotalUncert}{\SI{1.9}{\hertz}}
%
% This from the excel Scalar_Tensor shifts.xlsx, #m1.0-107-ge663bb:
\newcommand{\nonOrthogScalarShift}{\SI{106}{\hertz}}
\begin{figure}[tbh]
	\centering
	\includegraphics[width=\textwidth]{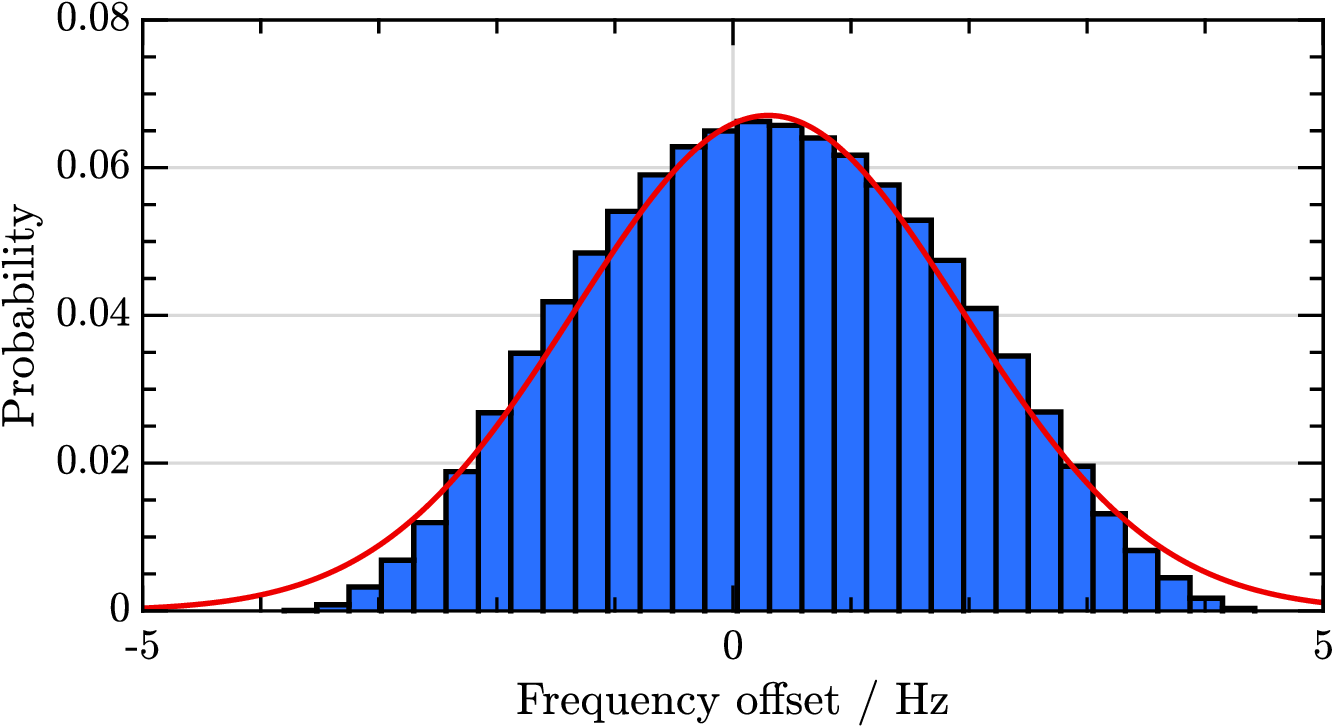}
	\caption{Range of errors possible for the tensor component of the shift to the E2 transition when averaged over the three orthogonal fields used in \cref{sec:polarizability_ratios}. For this run of the E2 scalar:tensor ratio dataset the magnetic field uncertainty was found to be $\sigma_B = \nonOrthogBFieldUncert$ on a field of \nonOrthogBField{}.
	The normal distribution least-squares fit to this range has a mean of \nonOrthogOffset{} and a 1-sigma range of \nonOrthogRange{}. The non-zero mean is due to granularity in our control of current in our trap's Helmholtz coils. For comparison, the scalar shift in this experimental configuration was \nonOrthogScalarShift{}.}
	\label{fig:non-orthogonal_fields_offset_histogram}
\end{figure}

The effect of a small additional background field would be to modify the
magnitudes and directions of all applied fields in the orthogonal set of 3
fields used in the experiment.  This would lead to a non-orthogonal set of ion
orientations, resulting in errors in the tensor shifts in \cref{eq:shift} and
an incomplete cancellation of these shifts when averaging over them to
determine the scalar shift.

Simulations were performed to evaluate the range of
frequency offsets possible due to incomplete cancellation of the tensor
component. This was done by computing a grid of all possible fields for the three
sets of currents used during each experiment, selected to be compatible with
the expected field to within the maximum offset observed in each dataset, $\sigma_B$. For each of these
candidate sets of fields, the average frequency shift was computed. 

For the ideal case of perfectly orthogonal fields this average should be zero,
and our analysis in \cref{tab:ratios} uses this fact to extract the scalar
component from the 3 measured \smh{} induced shifts.
\Cref{fig:non-orthogonal_fields_offset_histogram} shows the range of offsets possible when
calculating the scalar component of the E2 transition's polarizability-related
shift. The non-zero mean offset shown here is due to the limited resolution of our
current sources powering the compensation coils: the currents used produced
slightly non-orthogonal fields even without the addition of field noise. To be
conservative in our estimate of the effect of magnetic field
noise, we add the values of the mean offset and the 1-sigma width of the
distribution of offsets (\nonOrthogTotalUncert{} for this dataset) to produce
the final magnetic field-related uncertainty for the scalar shift evaluated
for this transition with this set of fields. For all ratio measurements, the scalar 
shift and uncertainty were evaluated in this way and repeated in an independent set
of orthogonal fields. As a consistency check, the ratio measurements on three occasions
and the mean values are reported in this paper. This simulation was also used to evaluate
errors on the tensor components of the relevant shifts using the same method. 

\FloatBarrier

\section*{References}

\bibliography{library}{}

\listoftodos

\end{document}